\documentclass[twocolumn, prl, aps, floats]{revtex4}

\DeclareMathAlphabet{\mathpzc}{OT1}{pzc}{m}{it}

\newcommand{\Nm}{\mathpzc{N}_{1}}
\newcommand{\nm}{\mathpzc{n}_{1}}
\newcommand{\sm}{\mathpzc{s}_{1}}

\usepackage{amsmath, amssymb, bbm, bm, color, epsf, epsfig, float, graphicx, ragged2e, relsize, setspace, yfonts}

\begin{document}
\begin{small}

\title{\flushleft
{\Large\rm Serendipity and strategy in rapid innovation} \\
{\rm\footnotesize T. M. A. Fink$^{* \dag}$, M.\ Reeves$^{\ddag}$, R. Palma$^{\ddag}$ and R.\ S.\ Farr$^{\dag}$} \\
\vspace{-3pt}
{\rm \footnotesize $^{\dag}$London Institute for Mathematical Sciences, Mayfair, London W1K 2XF, UK} \\
\vspace{-3pt}
{\rm \footnotesize $^{*}$Centre National de la Recherche Scientifique, Paris, France} \\
\vspace{-3pt}
{\rm \footnotesize $^{\ddag}$BCG Henderson Institute, The Boston Consulting Group, New York, USA} \\
\vspace{-3pt}
\mbox{}
\vspace{-30pt} 
\mbox{}
}


\maketitle


\begin{figure}[b]\setlength{\hfuzz}{1.1\columnwidth}
\begin{minipage}{\textwidth}
\includegraphics[width=1\textwidth]{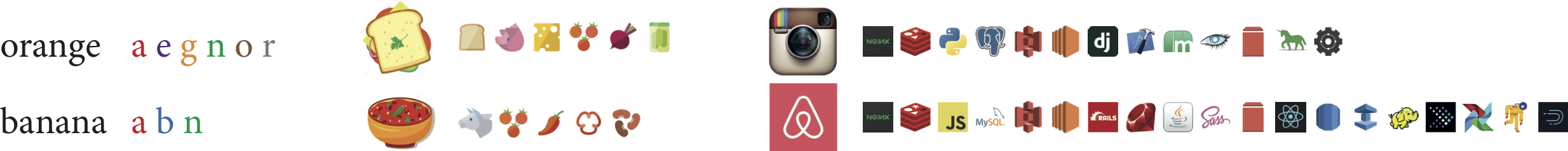}
\includegraphics[width=1\textwidth]{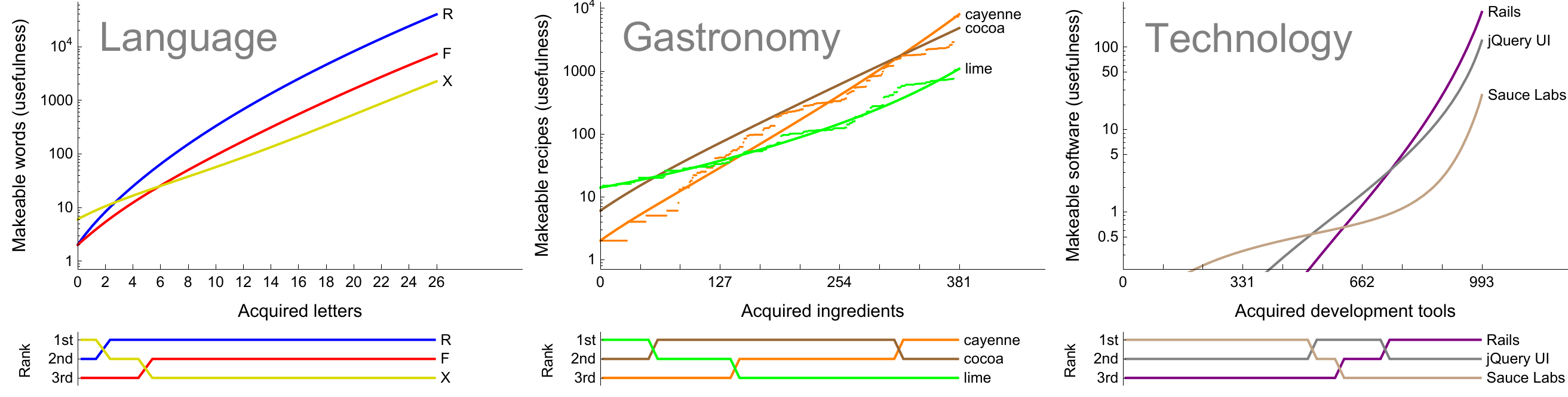}
\vspace{-15pt}
\caption{\footnotesize
\emph{Products, components and usefulness.}
({\bf Top})
We studied products and components from three sectors.
In language, 		the products are 79,258 English words 																and the components are the 26 letters. 
In gastronomy, 		the products are 56,498 recipes from the databases allrecipes.com, epicurious.com, and menupan.com \cite{ahnert14}	and the components are 381 ingredients.
In technology, 		the products are 1158 software products catalogued by stackshare.io										and the components are 993 development tools used to make them. 
({\bf Bottom})
The usefulness of a component is the number of products we can make that contain it. 
We find that the relative usefulness of a component depends on how many other components have already been acquired.
For each sector, we show the usefulness of three typical components: averaged at each stage over all possible choices of the other 
acquired components and---for gastronomy---for a particular random order of component acquisition (points). 
}
\end{minipage}
\label{products}
\end{figure}
\noindent
Innovation is to organizations what evolution is to organisms: it is how organisations adapt to changes in the environment and improve  \cite{erwin04}.
Governments, institutions and firms that innovate are more likely to prosper and stand the test of time;
those that fail to do so fall behind their competitors and succumb to market and environmental change \cite{reeves,weiss14}.
Yet despite steady advances in our understanding of evolution, what drives innovation remains elusive \cite{erwin04,farmer}.
On the one hand, organizations invest heavily in systematic strategies to drive innovation \cite{pietronero12,noorden15,drucker02,sood10}.
On the other, historical analysis and individual experience suggest that serendipity plays a significant role in the discovery process \cite{rosenman,johansson12,isaacson14}.
To unify these two perspectives, we analyzed the mathematics of innovation as a search process for viable designs across a universe of building blocks. 
We then tested our insights using historical data from language, gastronomy and technology. 
By measuring the number of makeable designs as we acquire more components,
we observed that the relative usefulness of different components is not fixed, but cross each other over time.
When these crossovers are unanticipated, they appear to be the result of serendipity. 
But when we can predict crossovers ahead of time, they offer an opportunity to strategically increase the growth of our product space.
Thus we find that the serendipitous and strategic visions of innovation can be viewed as different manifestations of the same thing: the changing importance of component building blocks over time.
\begin{figure}[b]
\includegraphics[width=0.77\columnwidth,height=1.07\columnwidth]{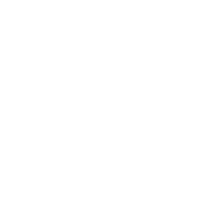}
\vspace{-15pt}
\label{products}
\end{figure}
\\ \indent
\emph{Lego game.}
Let's illustrate the idea using Lego bricks. 
Think back to your childhood days. 
You're in a room with two friends Bob and Alice, playing with a big box of Lego bricks---say, a fire station set. 
All three of you have the same goal: to build as many new toys as possible. 
As you continue to play, each of you searches through the box and chooses those bricks that you believe will help you reach this goal. 
Let's now suppose each player approaches this differently. 
Your approach is to follow your gut, arbitrarily selecting bricks that look intriguing.
Alice uses what we call a short-sighted strategy, carefully picking Lego men and their firefighting hats to immediately make simple toys.
Meanwhile, Bob chooses pieces such as axels, wheels, and small base plates that he noticed are common in more complex models, 
even though he is not able to use them straightaway to produce new toys. 
We call this a far-sighted strategy.
\\ \indent
\emph{Who wins.}
At the end of the day, who will have innovated the most? 
That is, who will have built the most new toys? 
We find that, in the beginning, Alice will lead the way, surging ahead with her impatient strategy. 
But as the game progresses, fate will appear to shift. 
Bob's early moves will begin to look serendipitous when he is able to assemble a complex fire truck from his choice of initially useless axels and wheels. 
It will seem that he was lucky, but we will soon see that he effectively created his own serendipity. 
What about you? 
Picking components on a hunch, you will have built the fewest toys. 
Your friends had an information-enabled strategy, while you relied on chance. 
\\ \indent
\emph{Spectrum of strategies.}
What can we learn from this? If innovation is a search process, then your component choices today matter greatly in terms of the options they will open up to you tomorrow. Do you pick components that quickly form simple products and give you a return now, or do you choose those components that give you a higher future option value? 
By understanding innovation as a search for designs across a universe of components, we made a surprising discovery. 
Information about the unfolding process of innovation can be used to form an advantageous innovation strategy.
But there is no one superior strategy.
As we shall see, the optimal strategy depends on time---how far along the innovation process we have advanced---and the sector---some sectors contain more opportunities for strategic advantage than others.
\\ \indent
\emph{Components and products.}
Just like the Lego toys are made up of distinct kinds of bricks, we take products to be made up of distinct components.
A component can be an object, like a touch screen, 
but it can also be a skill, like using Python, or a routine, like customer registration. 
Only certain combinations of components form products, according to some predetermined universal recipe book of products.
Examples of products and the components used to make them are shown in  Fig. 1.
Now suppose that we possess a basket of distinct components, which we can combine in different ways to make products. 
We have more than enough copies of each component for our needs, so we do not have to worry about running out.
There are $N$ possible component types in total, but at any given stage $n$ we only have $n$ of these $N$ possible building blocks.
At every stage, we pick a new type of component to add to our basket. 
\begin{figure}[b!]
\includegraphics[width=1.01\columnwidth]{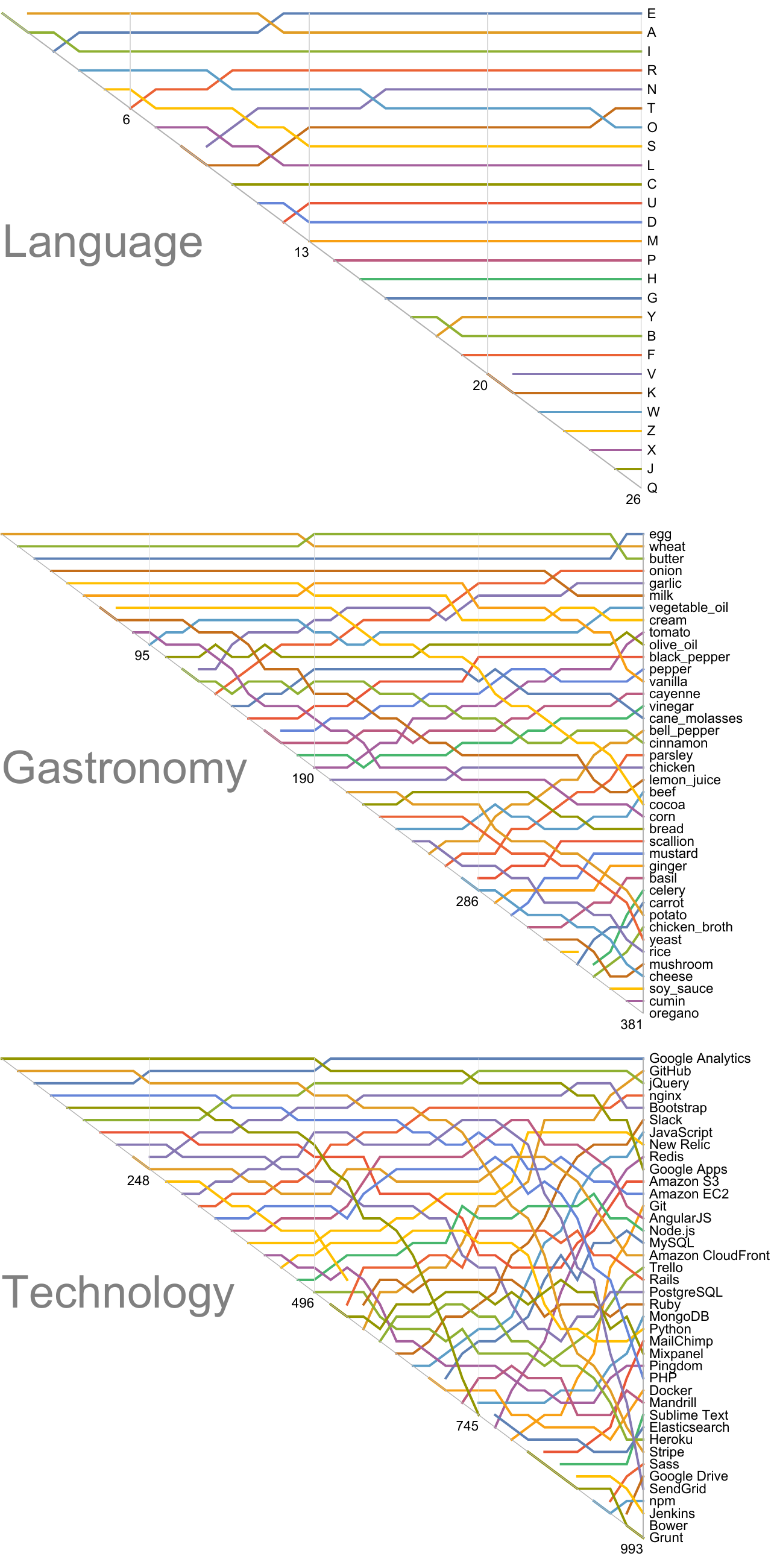}
\vspace{-88pt}
\caption{\footnotesize 
\emph{Crossovers}.
The relative \\
usefulness of different components \\
changes as the number of components \\
we possess increases. For example, if you \\
are only allowed six letters, the ones that show \\
up in the most words are \emph{a}, \emph{e}, \emph{i}, \emph{o}, \emph{s}, \emph{r}.
For gastro- \\
nomy and technology, for clarity we only show the \\
40 components most useful when 
we have all $N$ components.
A pure short-sighted strategy acquires components in the order that they intersect the diagonal; whereas 
a pure far-sighted strategy acquires them in the order that they intersect a vertical.
If there are no crossovers, the strategies are the same.
}
\label{combined}
\end{figure}
\begin{figure}[b!]
\includegraphics[width=0.96\columnwidth]{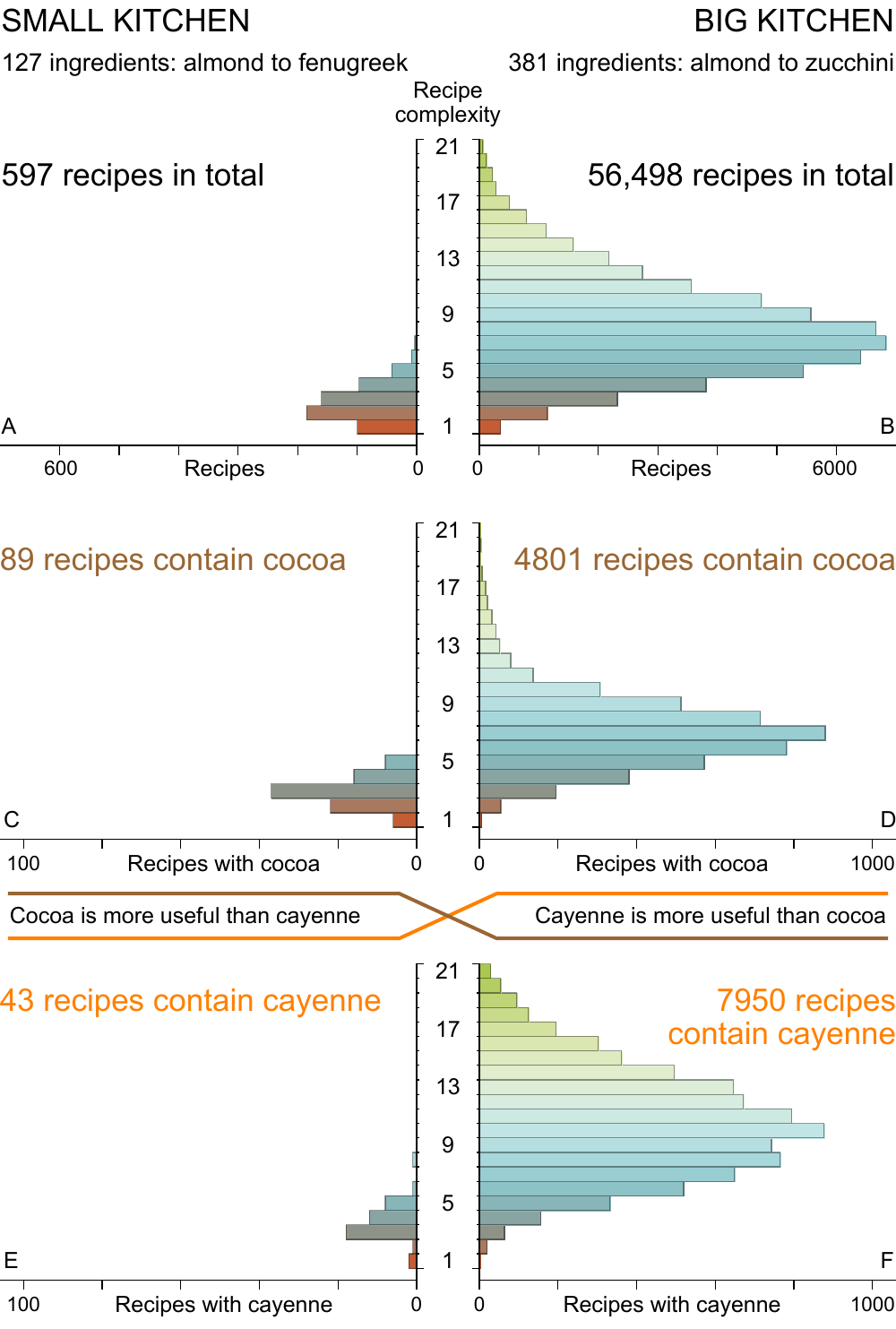}
\vspace{-10pt}
\caption{\footnotesize
\emph{Why crossovers happen.}
On the right is a big kitchen with 381 ingredients.
On the left is a small kitchen with one-third as many ingredients.
In the big kitchen ({\bf B}), we can make a total of 56,498 recipes. 
Each bar counts recipes with the same number of ingredients (complexity).
When we move to the smaller kitchen ({\bf A}), the number of makable recipes shrinks dramatically to 597, or 1.0\%.
But this reduction is far from uniform across different bars. 
Higher bars shrink more, on average by an extra factor of 3 with each bar.
Thus the number of recipes of complexity one (first bar) shrinks about 3-fold;
the number of complexity two (second bar) 9-fold, and so on.
Of all the recipes in the big kitchen,
4801 contain cocoa  ({\bf D}) and
7950 contain cayenne ({\bf F}). 
The cayenne recipes tend to be more complex, containing on average 10.6 ingredients, whereas the cocoa recipes are simpler, averaging 7.2 ingredients.
Because higher bars suffer stronger reduction, overall fewer cayenne recipes (0.5\%) survive in the smaller kitchen ({\bf E}) than cocoa recipes (1.8\%) ({\bf C}).
Thus cayenne is more useful in the big kitchen, but cocoa is more useful in the small kitchen.
}
\label{strategy}
\end{figure}
\\ \indent
\emph{Usefulness.}
The usefulness of a component is the number of products we can make that contain it \cite{products}.
In other words, the usefulness $u_\alpha$ of some component $\alpha$ is how many more products we can make with $\alpha$ in our basket than without $\alpha$ in our basket.
As we gather more components, $u_\alpha$ increases or stays the same; it cannot decrease. 
We write $u_\alpha(\mathpzc{n})$ to indicate this dependence on $n$: $u_\alpha(\mathpzc{n})$ is the usefulness of $\alpha$ given possession of $\alpha$ and $n-1$ other components, 
the combined set of components being $\mathpzc{n}$.
Averaging over all choices of the $n-1$ other components from the $N-1$ that are possible gives the mean usefulness, $\overline{u}_\alpha(n)$.
\\ \indent
\emph{Usefulness experiment.}
To measure the usefulness of different components as the innovation process unfolds and we acquire more components, we did the following experiments.
Using data from each of our three sectors, 
we put a given component $\alpha$ into an empty basket, and then added, one component at a time, 
the remaining $N-1$ other components, measuring the usefulness of $\alpha$ at every step.
We averaged $u_\alpha(\mathpzc{n})$ over all possible orders in which to add the $N-1$ components to obtain $\overline{u}_\alpha(n)$. 
(We explain how in SI B.)
We repeated this process for all of the components $\alpha$.
Typical results from these experiments are shown in Fig. 1.
We find that the mean usefulnesses of different components \emph{cross each other} as the number of components in our basket increases.
As Fig. 1 shows for gastronomy, this is true for both the average over all possible orderings of components (lines)
as well as a specific random ordering (points). 
\\ \indent
\emph{Bumps charts.}
To visualise the relative usefulness of components over time, for each sector we created its ``bumps chart" (Fig. 2).
These show the rank order of mean usefulness at every stage of the innovation process.
We see that the crossovers in Fig. 1 are commonplace, but that some sectors contain more crossovers than others.
There are few crossings in language, some in gastronomy and many in technology. 
This means, for example, that the most useful letters for making words in Scrabble (a basket of seven letters) 
are nearly the same as the most useful letters for making words with a full basket (26 letters);
the key ingredients in a small kitchen (20 ingredients) are moderately different from those in a big one (80 ingredients);
the most-used development skills for a young software firm (experience with 40 tools) are significantly different from those for an advanced one (160 tools).
We call components that do not cross in time \emph{isochronic}, like the letters; and those that do \emph{anisochronic}, like the tools.
\begin{figure}[b!]\setlength{\hfuzz}{1.1\columnwidth}
\begin{minipage}{\textwidth}
\includegraphics[width=1\textwidth]{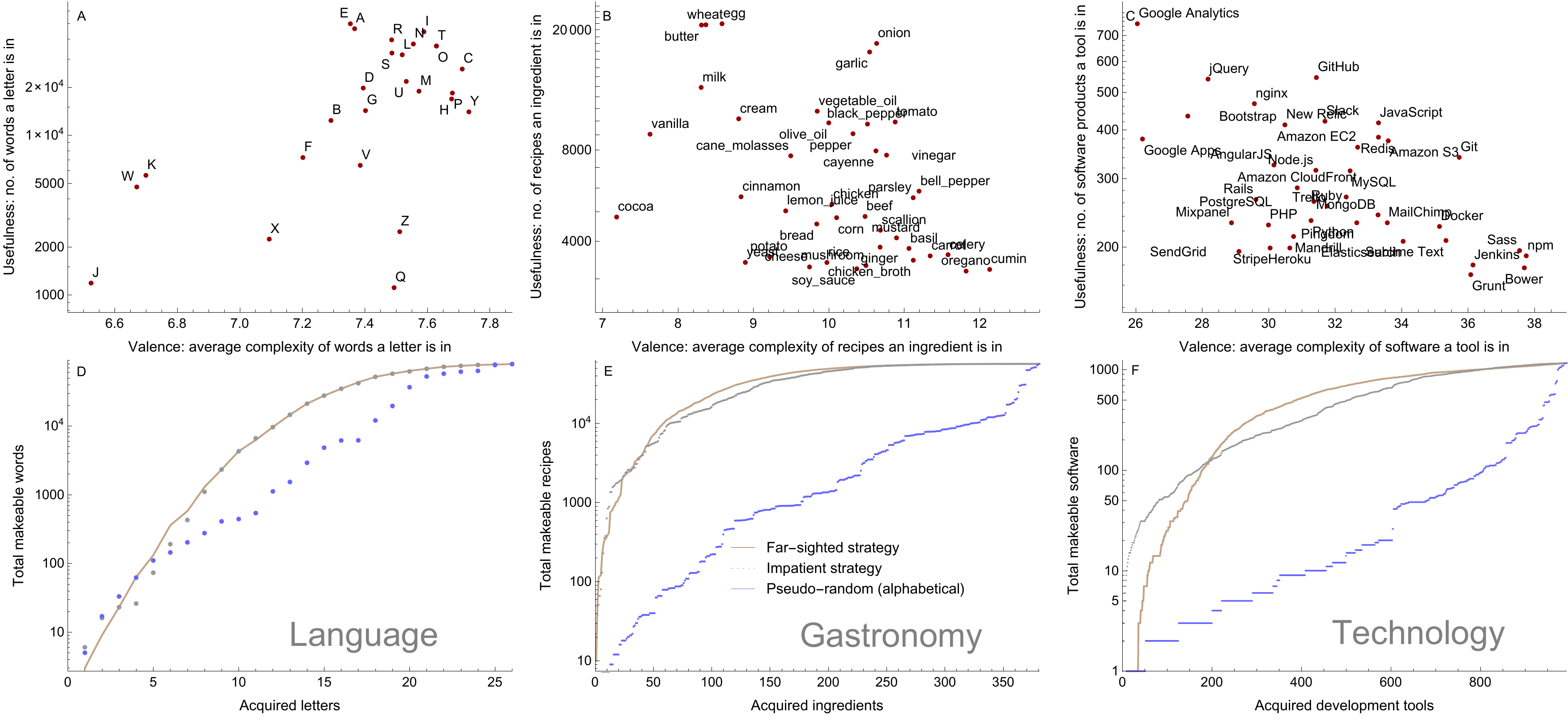}
\vspace{-18pt}
\caption{\footnotesize
({\bf ABC}) Scatter plots of component usefulness versus component valence for our three sectors.
For gastronomy and technology, we only show the top 40 components; the complete set is in SI Fig. 5.
({\bf DEF}) Both the short-sighted and far-sighted strategies beat a typical random component ordering (here alphabetical), 
but they diverge from each other only insofar that there are crossings in the bumps charts.
}
\end{minipage}
\label{scatter}
\end{figure}
\\ \indent
\emph{Why crossovers happen.}
To understand why crossovers happen, let's have a closer look at how the mean usefulness increases for a single component (Fig. 3).
To make a product of complexity $s$, we must possess all $s$ of its distinct components.
So making a complex product is harder than making a simple one,
because there are more ways that we might be missing a necessary component.
We therefore group together the products we can make containing $\alpha$ according to their complexity.
That is, the usefulness $u_\alpha(\mathpzc{n},s)$ of component $\alpha$ is how many more products of complexity $s$ we can make with $\alpha$ in our basket than without $\alpha$ in our basket.
Summing $u_\alpha(\mathpzc{n},s)$ over $s$ gives $u_\alpha(\mathpzc{n})$.
The advantage of this refined grouping is that, by understanding the behaviour of $\overline{u}_\alpha(n,s)$, we can understand the more difficult $\overline{u}_\alpha(n)$.
Our key result, which we prove in SI B, is that $\overline{u}_\alpha(n,s)/n^{s-1}$ is constant over all stages of the innovation process. 
In other words, for two stages $n$ and $n'$,
\begin{equation}
\overline{u}_{\alpha}(n',s) \simeq \overline{u}_{\alpha}(n,s)  (n'/n)^{s-1}.
\label{FF}
\end{equation}
This tells us that
the number of products containing $\alpha$ of complexity $s$ grows much faster for higher complexities than for lower complexities.
Early on, $\overline{u}_{\alpha}(n,s)$ will tend to be small for higher complexities, but depending on how far ahead we look, the bigger growth rate can more than compensate for this,
as we see in Fig. 3.
Summing eq. (\ref{FF}) over size $s$, we find
\begin{equation}
\overline{u}_\alpha(n') \simeq \overline{u}_\alpha({n},1) + \overline{u}_\alpha({n},2) \, x + \overline{u}_\alpha({n},3) \, x^2 + \ldots,
\label{GG}
\end{equation}
where $x=n'/n$.
The growth of the mean usefulness of $\alpha$ strongly depends on the complexity of products containing $\alpha$.
\\ \indent
\begin{figure}[b]
\includegraphics[width=0.76\columnwidth,height=1.07\columnwidth]{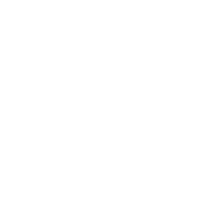}
\end{figure}
\emph{Valence.}
So far we have only characterised a component by its usefulness: the number of products we can make that contain it. 
Now we introduce another way of describing a component: the average complexity of the products it appears in.  We call this the \emph{valence}. 
The valence $v_\alpha$ of component $\alpha$ is the average complexity of the products it appears in at stage $N$, when we have all $N$ components.
Think of the valence as the typical number of co-stars a component performs with, plus one.
We show the usefulness and valence for each of the components in our three sectors in Fig. 4ABC.
More valent components are unlikely to be useful until we possess a lot of other components, so that we have a good chance of hitting upon the ones they need. 
These are the wheels and axels in our Lego set.
On the other hand, less valent components are likely to boost our product space early on, when we have acquired fewer components.
These are the Lego men and their firefighting hats.
This insight suggests that more valent components will tend to rise in relative usefulness, and less valent components fall.
This is verified in our experiments: components on the right of the plots in Fig. 4ABC tend to rise in the bumps charts in Fig. 2, such as onion, tomato, Javascript and Git;
whereas components on the left tend to fall, like cocoa, vanilla, Google Apps and SendGrid.
%
\\ \indent
\emph{Interpreting crossovers.}
A crossover in the usefulness of components means that the things that matter most today are not the same as the things that will matter most tomorrow.
How we interpret crossovers in practice depends on whether they are unanticipated, and take us by surprise, or anticipated, and can be planned for and exploited.
When they are unanticipated, beneficial crossovers can seem to be serendipitous.
But when they can be anticipated, crossovers provide an opportunity to strategically increase the growth of our product space.
To harness this opportunity, we turn to forecasting component crossovers using the complexity of products containing them.
\\ \indent
\emph{Short-sighted strategy.}
To maximise the size of our product space when crossovers are unanticipated,
the optimal approach is to acquire, at each stage, the component that is most useful from the ones that are remaining. Think of this as a ``greedy'' approach.
It has a geometric interpretation: it is equivalent to acquiring the components that intersect the diagonals in Fig. 2.
At every stage we lock in to a specific component, unaware of the future implications of the choices we make. 
A component poorly picked is an opportunity lost.
\\ \indent
\emph{Far-sighted strategy.}
Using only information about the products we can already make with our existing components, however, 
we can forecast the usefulness of our components into the future.
Eq. (\ref{GG}) shows us how, and we give an example in SI C.
Here 	the optimal approach		is to acquire 				the component that will be most useful at some later stage $n'$.
This also has a geometric interpretation: it is equivalent to acquiring the components that intersect a vertical at $n'$ in Fig. 2,
and thus depends on how far into the future we forecast.
\\ \indent
\emph{Strategy comparison.}
A short-sighted strategy considers only the usefulness $u_\alpha$, 
whereas a far-sighted strategy considers both the usefulness $u_\alpha$ and the valence $v_\alpha$.
Short-sighted 	maximises what a potential new component 	can do for us \emph{now}, whereas
far-sighted 	maximises what it 						could do for us \emph{later}.
Depending on our desire for short-term gain versus long-term growth, we have a spectrum of strategies dependent on $n'$.
A pure short-sighted strategy ($n'=n$) and a pure far-sighted strategy ($n'=N$) are compared in Fig. 4DEF.
Like the Lego approaches of Bob and Alice, both strategies beat acquiring components in a random order.
As our theory predicts, the extent to which the two strategies differ from each other increases with the number of crossovers.
For language, 		they are nearly identical, because there are hardly any crossovers.
For gastronomy, 	short-sighted has a two-fold advantage at first, 			but later far-sighted wins by a factor of two.
For technology, 	short-sighted surges ahead by an order of magnitude, 	but later far-sighted is dominant.
\\ \indent
\emph{Serendipity and strategy.}
Our research helps resolve the tension between a strategic approach to innovation, 
which views innovation as a rational process which can be measured and prescribed \cite{drucker02,sood10,weiss14,farmer};
and a belief in serendipity and the intuition of extraordinary individuals \cite{rosenman,johansson12,isaacson14}.
A strategic approach is seen in firms like P\&G and Unilever, 
which use process manuals and consumer research to maintain a reliable innovation factory \cite{brown},
and Zara, which systematically scales new products up and down based on real-time sales data.
In scientific discovery, ``traditional scientific training and thinking favor logic and predictability over chance'' \cite{rosenman}.
If discoveries are actually made in the way that scientific publications suggest, the path to invention is a step-by-step, rational process.
On the other hand, a serendipitous approach is seen in firms like Apple, which is notoriously opposed to making innovation choices based on incremental consumer demands,
and Tesla, which has invested for years in their vision of long-distance electric cars \cite{bullis}.
In science, many of the most important discoveries have serendipitous origins, in contrast to their published step-by-step write-ups, 
such as penicillin, heparin, X-rays and nitrous oxide  \cite{rosenman}.
The role of vision and intuition tend to be under-reported:
a study of 33 major discoveries in biochemistry ``in which serendipity played a crucial role'' concluded that ``when it comes to `chance' factors, few scientists `tell it like it was'" \cite{tria,comroe}.
\\ \indent
\emph{Serendipity}.
Writing about the \emph{The Three Princes of Serendip}, Horace Walpole records that the princes
``were always making discoveries, by accidents and sagacity, of things they were not in quest of''.
Serendipity is the fortunate development of events, and many organizations and researchers stress its importance \cite{rosenman,johansson12}. 
Crossovers in component usefulness help us see why.
Components which depend on the presence of many others can be of little benefit early on.
But as the innovation process unfolds and the acquired components pay off, the results will seem serendipitous, 
because a number of previously low-value components become invaluable. 
Thus, what appears as serendipity is not happenstance but the delayed fruition of components reliant on the presence of others.
After the acquisition of enough other components, these components flourish.
For example, the initially useless axels and wheels were later found to be invaluable to building many new toys.
In a similar way, the low value attributed to Flemming's initial identification of lysosome was later revised to high value in the years leading to the discovery of penicillin,
when other needed components emerged, such as sulfa drugs which showed that safe antiseptics are possible \cite{rosenman}.
Interestingly, the word ``serendipity'' does not have an antonym.
But as our bumps charts show, for every beneficial shift in a crossover, there is a detrimental one.
Each opportunity for serendipity goes hand-in-hand with a chance for \emph{anti-serendipity}:
the acquisition of components useful now but less useful later.
Avoiding these over-valued components is as important as acquiring under-valued ones to securing a large future product space.
\\ \indent
\emph{Strategy.}
Our research shows that the most important components---materials, skills and routines---when an organization is less developed tend to be different from when it is more developed.
Instead, the relative usefulness of components can change over time, in a statistically repeatable way.
Recognising how an organization's priorities depend on its maturity enable it to balance short-term gain with long-term growth. 
For example, our insights provide a  framework for understanding the poverty trap.
When a less-developed country imitates a more-developed country by acquiring similar production capabilities \cite{pietronero12}, it is unable to quickly reap the rewards of its investment,
because it does not have in place enough other needed capabilities. 
This in turn prevents it from further investment in those needed components.
Our analysis gives quantitative backing to the ``lean start-up'' approach to building companies and launching products \cite{lean}.
Start-ups are wise to employ a short-sighted strategy and release a minimum viable product.
Without the resources to sustain a far-sighted approach, they need to quickly bring a simple product to market. 
On the other hand, firms that can weather an initial drought will see their sacrifice more than paid off  when their far-sighted approach kicks in.
By tracking how potential new components combine with existing ones, organisations can develop an information-advantaged strategy to adopt the right components at the right time.
In this way they can create their own serendipity, rather than relying on intuition and chance.

\noindent

{\large Online supplementary information (SI)} 
\\ \mbox{} \\
{\bf A.\ Data} \\
Our three data sets---described in Fig.\ 1---were obtained as follows.
In language, our list of 79,258 common English words is from the built-in WordList library in Mathematica 10. 
Of the 84,923 KnownWords, we only considered those made from the 26 letters a--z, ignoring case: we excluded words containing a hyphen, space, etc.
In gastronomy, the 56,498 recipes can be found in the supplementary material in \cite{ahnert14}.
In technology, the 1158 software products and the development tools used to make them can be found at the site stackshare.io.
\\ \mbox{} \\
\noindent
{\bf B.\ Proof of components invariant} \\
Let $\alpha$ be some component.
Let $\Nm$ be the set of $N-1$ other possible components not including $\alpha$, 
$\nm$ be a subset of $n-1$ components chosen from $\Nm$, and
$\sm$ be a subset of $s-1$ components chosen from $\nm$.
The usefulness $u_{\alpha}(\mathpzc {n},s)$ is how many more products of complexity $s$ that we can make from the components $\nm$ together with $\alpha$, 
than from the components $\nm$ alone:
\begin{equation*}
u_{\alpha}(\mathpzc {n},s) = \sum_{\mathpzc{\sm} \subseteq \mathpzc{\nm}} {\rm prod}(\alpha \cap \sm)  - {\rm prod}(\sm),
\end{equation*}
where prod$(\alpha \cap \sm)$ takes the value 
0 if the combination of components $\alpha \, \cap \, \sm$ forms no products of complexity $s$ and
1 if $\alpha \cap \sm$ forms one product of complexity $s$.
(Occasionally, the same combination of components $\alpha \cap \sm$ forms multiple products: for example, beef, butter and onion together form two distinct recipes of length three.
In such cases, prod$(\alpha \cap \sm)$ takes the value 2 if $\alpha \, \cap \, \sm$ forms two products, and so on.)
The expected 		usefulness of component $\alpha$, 	$\overline{u}_{\alpha}(n,s)$, 	is the average of	$u_{\alpha}(\mathpzc{n},s)$	over all subsets $\nm \subseteq \Nm$;
there are ${N-1 \choose n-1}$ such subsets.
Therefore
\begin{eqnarray}
\overline{u}_\alpha(n,s) 	&\!=\!& {1}\big/{\textstyle{N-1 \choose n-1}} \sum_{\nm \subseteq \Nm} u_\alpha(\mathpzc{n},s) \nonumber \\
				 	&\!=\!& {1}\big/{\textstyle{N-1 \choose n-1}} \sum_{\nm \subseteq \Nm}  \,\, \sum_{\sm \subseteq \nm}  {\rm prod}(\alpha \cap \sm)  - {\rm prod}(\sm) \nonumber.
\label{PP}
\end{eqnarray}
Consider some particular combination of components $\sm'$.
The double sum above will count $\sm'$ once if $s=n$, but multiple times if $s < n$, 
because $\sm'$ will belong to multiple sets $\nm$.  
How many?
In any set $\nm$ that contains $\sm$, there are $n-s$ free elements to choose, from $N-s$ other components.
Therefore the double sum will count every combination $\sm$ a total of ${N-s \choose n-s}$ times, and
\begin{eqnarray*}
\overline{u}_{\alpha}(n,s) 		&=& 						{\textstyle {N-s \choose n-s}} \big/ {\textstyle{N-1 \choose n-1}}
												\sum_{\mathpzc{\sm} \subseteq \mathpzc{\Nm}} {\rm prod}(\alpha \cap \sm)  - {\rm prod}(\sm) \\
						&=&	\textstyle N/n \, {{n \choose s}}\big/{{N \choose s}}  \, \overline{u}_{\alpha}(\mathpzc{N},s).
\end{eqnarray*}
The same must be true when we replace $n$ by $n'$, and therefore
\begin{equation}
\textstyle {\overline{u}_{\alpha}(n,s)} \, n / \binom{n}{s} = {\overline{u}_{\alpha}(n',s)} \, n' / \binom{n'}{s}.
\label{exact}
\end{equation}
When the number of components is big compared to the product size ($n, n' \gg s$), 
we can approximate $\binom{n}{s}$ and $\binom{n'}{s}$ by $n^s$ and $n'^s$, and thus
\begin{equation*}
\textstyle {\overline{u}_{\alpha}(n,s)} / n^{s-1} \simeq {\overline{u}_{\alpha}(n',s)} / n'^{s-1}.
\end{equation*}
For simplicity, we use this approximation in the main manuscript, but we could just as well have used the exact expression  in eq. (\ref{exact}).
\\ \noindent \\
{\bf C.\ Forecasting crossovers in usefulness} \\
Here we show how we can forecast the usefulness of components at stage $n'$ from information we have at some earlier stage $n$,
where $n$ is the number of components we have acquired.
As in Fig. 3, we have a set $\mathpzc{k}$ of 127 ingredients in a small kitchen---almond to fenugreek---and
a set $\mathpzc{K}$ of 381 ingredients in a big kitchen---almond to zucchini.
\\ \indent
In the small kitchen, we can make a total of 597 recipes.
Of these 597 recipes, 43 contain cayenne, but they are not all equally complex.
Two of the 43 recipes contain one ingredient (namely, cayenne itself) and have complexity one;
one recipe contains two ingredients and has complexity two; 
18 contain three ingredients and have complexity three; 
and so on.
Similarly, 89 of the 597 recipes contain cocoa: 
six have complexity one;
22 have complexity two;
and so on.
Using eq. (\ref{GG}), we can write the mean usefulness of these two components as
\begin{eqnarray*}
\overline{u}_{\rm ca}(n' |\mathpzc{k}) &\simeq& 2 + x + 18 x^2 + 12 x^3 + 8 x^4 + x^5 + x^7  \quad {\rm and} \\
\overline{u}_{\rm co}(n' |\mathpzc{k}) &\simeq& 6 + 22 x + 37 x^2 + 16 x^3 + 8 x^4,
\end{eqnarray*}
where $x = n'/127$. 
As expected,
\begin{eqnarray*}
\overline{u}_{\rm ca}(n'|\mathpzc{k})\big\vert_{x=1} &=& 43 \quad {\rm and} \\
\overline{u}_{\rm co}(n'|\mathpzc{k})\big\vert_{x=1} &=& 89.
\end{eqnarray*}
\indent
In the big kitchen, we can make a total of 56,498 recipes.
Of these, 7950 contain cayenne and 4801 contain cocoa.
Again using eq. (\ref{GG}),
\begin{eqnarray*}
\overline{u}_{\rm ca}(n' |\mathpzc{K}) &\simeq& 2 + 19 x + 64 x^2 + \ldots + 2 x^{28} + 2 x^{30}  \quad {\rm and} \\
\overline{u}_{\rm co}(n' |\mathpzc{K}) &\simeq& 6 + 54 x + 195 x^2 + \ldots + 2 x^{20} + 3 x^{21}.
\end{eqnarray*}
where $x = n'/381$. As expected,
\begin{eqnarray*}
\overline{u}_{\rm ca}(n'|\mathpzc{K})\big\vert_{x=1} &=& 7950 \quad {\rm and} \\
\overline{u}_{\rm co}(n'|\mathpzc{K})\big\vert_{x=1} &=& 4801.
\end{eqnarray*}
\indent
So far, none of this is surprising.
The punchline is that we can estimate the usefulness of components in the big kitchen from what we know about our small kitchen.
To do so, we simply evaluate the small-kitchen polynomials at the big-kitchen stage:
\begin{eqnarray*}
\overline{u}_{\rm ca}(n'|\mathpzc{K})\big\vert_{n'=381} &\simeq& \overline{u}_{\rm ca}(n'|\mathpzc{k})\big\vert_{x=3} \simeq 3569 \quad {\rm and} \\
\overline{u}_{\rm co}(n'|\mathpzc{K})\big\vert_{n'=381} &\simeq& \overline{u}_{\rm co}(n'|\mathpzc{k})\big\vert_{x=3} \simeq 1485.
\end{eqnarray*}
In log terms---log usefulness being the natural unit of measure---these are accurate to within 11\% and 9\% of the true values. 
In particular, this predicts the crossover of cayenne and cocoa in Figure 3.
\begin{figure*}[b!]
\includegraphics[width=0.95\textwidth]{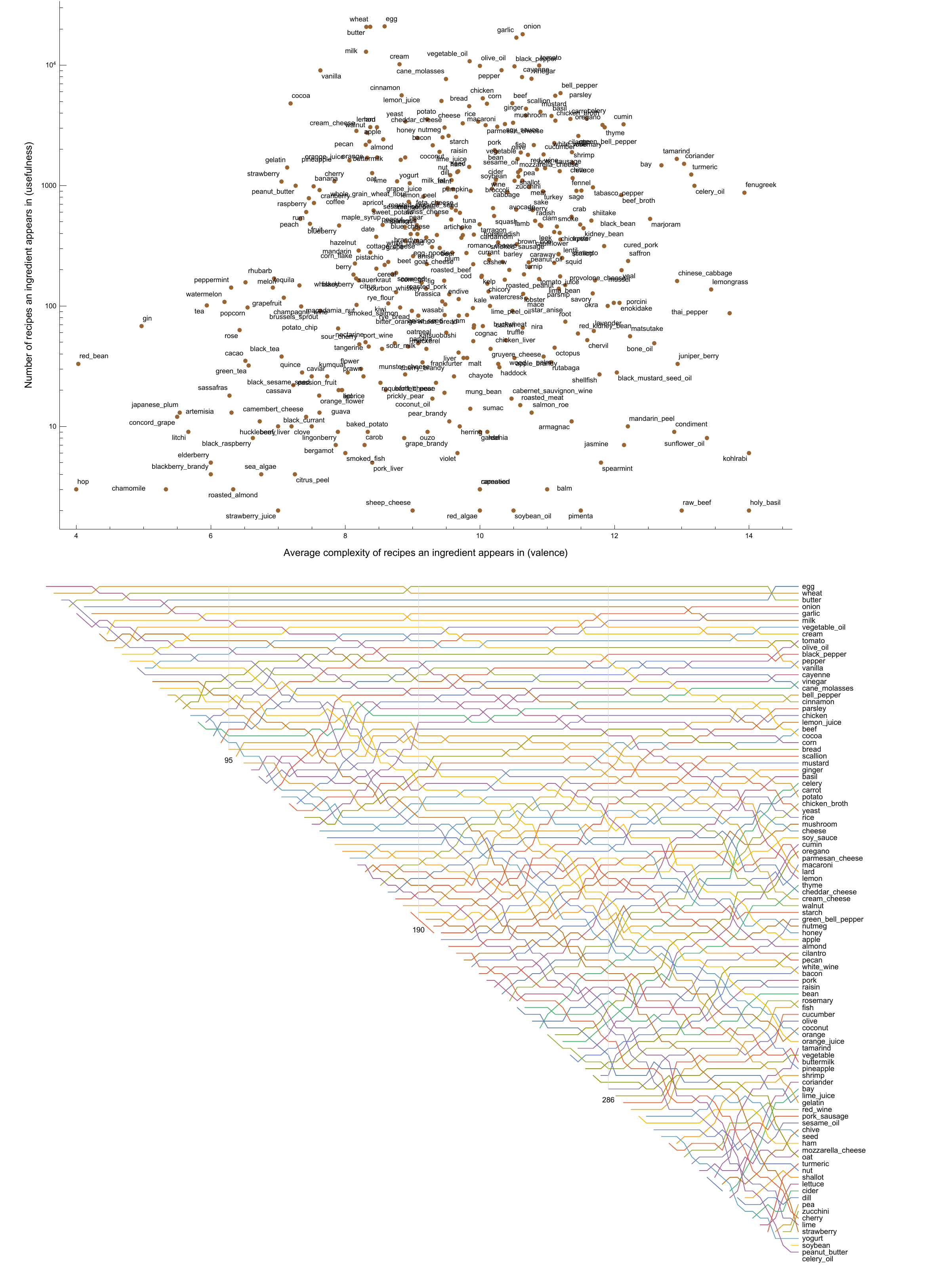}
\caption{\footnotesize 
({\bf Top}) The valence-usefulness scatter plot for all ingredients that are used in two or more recipes (365 of the 381 ingredients).
({\bf Bottom}) The relative usefulness of different ingredients as the number of ingredients we possess increases, for the 100 ingredients most useful when we have all 381 ingredients.
}
\label{strategy}
\end{figure*}
\begin{figure*}[b!]
\includegraphics[width=.95\textwidth]{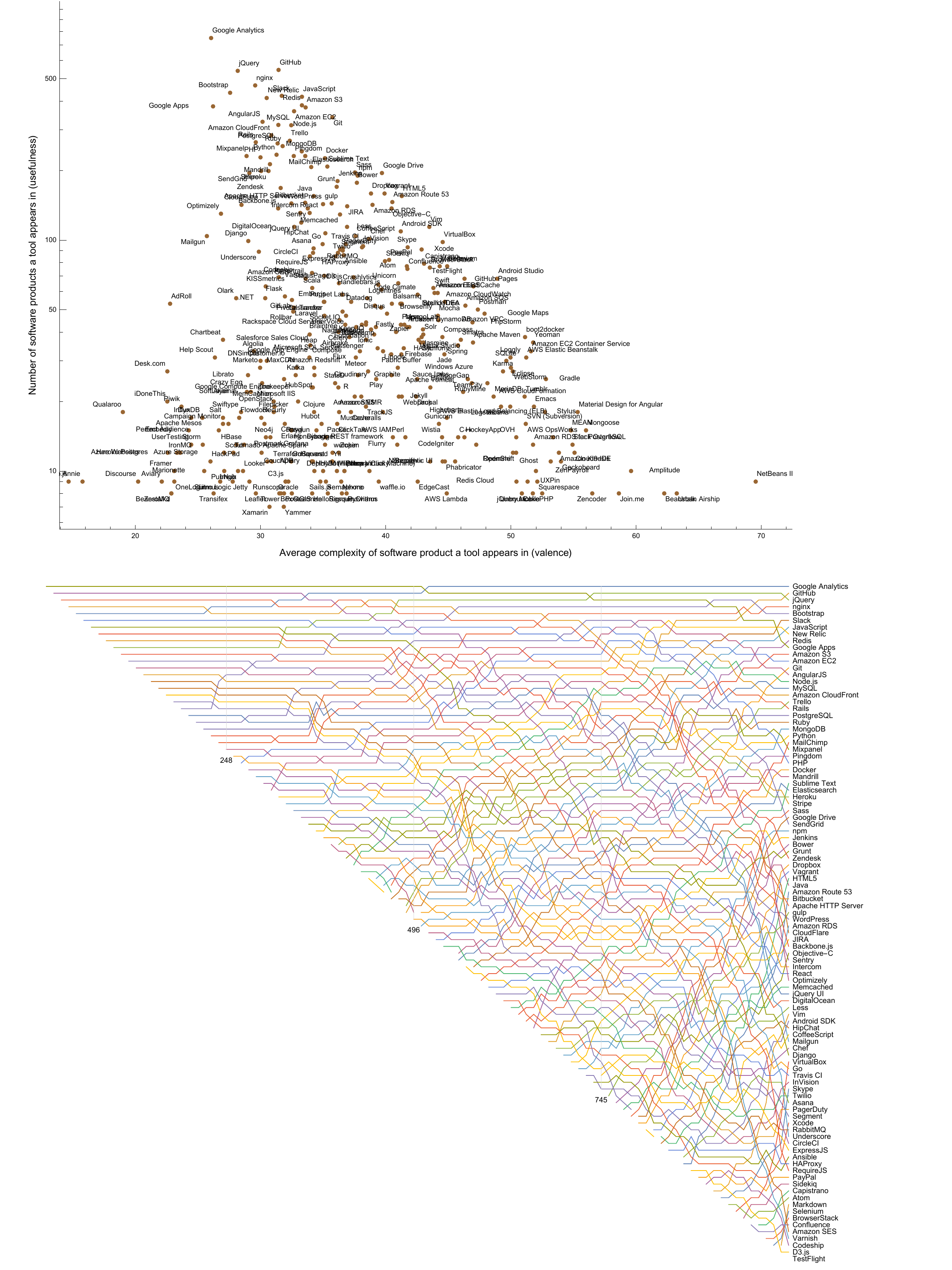}
\caption{\footnotesize 
({\bf Top}) The valence-usefulness scatter plot for the 365 technology tools most useful in making software products.
({\bf Bottom}) The relative usefulness of different tools as the number of tools we possess increases, for the 100 tools most useful when we have all 993 tools.
}
\label{strategy}
\end{figure*}


\end{small}

\end{document}